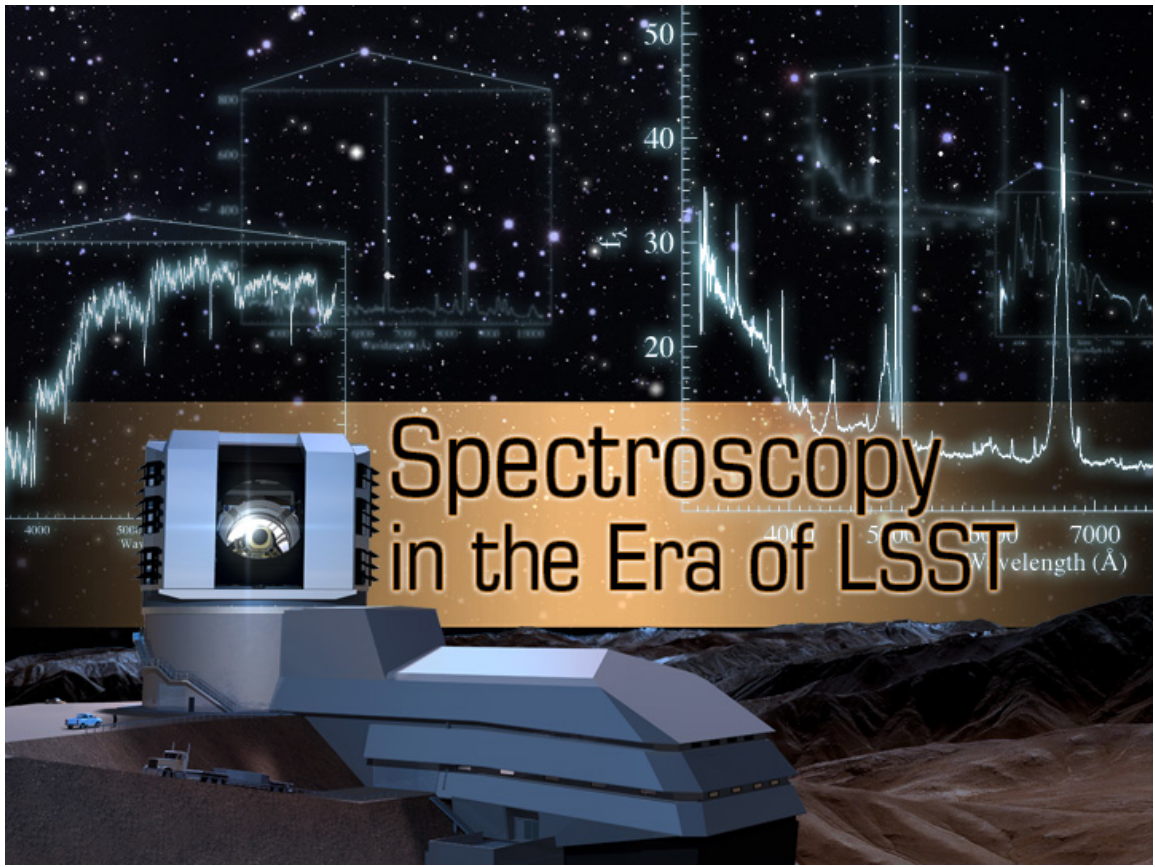

# Spectroscopy in the Era of LSST

Thomas Matheson (NOAO), Xiaohui Fan (Arizona), Richard Green (Arizona), Alan McConnachie (NRC-HIA), Jeff Newman (Pittsburgh), Knut Olsen (NOAO), Paula Szkody (Washington), & W. Michael Wood-Vasey (Pittsburgh)

## 1. Introduction

The Large Synoptic Survey Telescope (LSST) represents a significant change in the strategies employed by U.S. astronomers to study the Cosmos.  The Survey will image the available Southern sky every three to four days, switching among the six filters (*ugrizy*) over ten years to achieve depth and uniformity.  The four main scientific goals of the Project are: 1) characterize the nature of dark energy, 2) illuminate the structure of the Milky Way, 3) perform a census of the Solar System, and 4) discover transient and variable objects.  These goals can be achieved within the context of the project itself, but each of them can be greatly enhanced with the addition of a spectroscopic component. Moreover, spectroscopic follow-up will enable an even broader range of scientific use of LSST.

To begin the discussion of the spectroscopic capabilities desired for LSST follow-up, the National Optical Astronomy Observatory (NOAO) hosted a two-day workshop (April 11-12, 2013) in Tucson. There were over sixty attendees representing a wide range of US astronomical interests, including public and private universities, federal facilities, and international observatories. Plenary sessions gave an overview of the LSST project itself, as well as a broad vision of the astronomical landscape in 2020, near the start of the operations phase of LSST. Breakout sessions were organized by science topic. Each session had a facilitator to guide discussion and provide feedback to the entire workshop. The four areas were: time domain science, Galactic structure and stellar populations, galaxies and AGN, and dark energy and cosmology. There were not enough attendees with interest in the Solar System to convene a session on that topic.

The goal of the breakout sessions was to identify spectroscopic capabilities that each science area would like to have to follow-up of LSST discoveries. The focus was on science-driven needs, with brief science cases to be included. No specific budget constraints were applied, but attendees were reminded to be realistic in their expectations. The capabilities to consider were not just new or repurposed instruments. The participants in each session were also encouraged to explore new facilities, observing modes, time-allocation modes, and software infrastructure to assist with target selection and scheduling. In addition, they could also describe precursor experiments that might guide LSST follow-up. For all capabilities, the need for specificity was emphasized.

During the sessions, a table was devised as a way to characterize the specific needs for individual science cases (see Appendix 1 for examples from each breakout session). This included such information as the necessary spectral resolution, wavelength coverage, density of targets (single vs. multi-objects), depth (aperture), number of targets, and many others. Each session contributed examples to demonstrate the utility of such a table. This table is still a work in progress, but can serve as a framework for future discussions of spectroscopic capabilities.

Section 2 includes the reports from the breakout sessions. Section 3 synthesizes the capabilities from the breakout sessions and assesses the common desires.

**2. Reports from Breakout Sessions**

2.1 Time Domain (Paula Szkody facilitator)

The LSST project presents a tremendous challenge to astronomers interested in time-domain science. The volume and data rate will be unprecedented. Current time-domain projects (e.g., Palomar Transient Factory, Catalina Real-Time Transient

Survey, PAN-Starrs, La Silla-QUEST, Skymapper) could saturate available spectroscopic facilities now, and LSST will only add to the demand.

Prior to the start of the main LSST survey, a time domain ecosystem must be established and tested to enable the capability to sort rare and unusual phenomena from normal known variable objects. This will require precursor community programs (currently and including LSST itself during commissioning) using 2-8m class telescopes over small areas of the sky to identify normal transients, with a cadence from 15 seconds to 6 months and filters similar to those to be used by LSST. The light curves obtained need to be followed up with low resolution spectra for positive identification and training of classification schema and to set up an event broker (software infrastructure that can filter alerts). In addition, testing of rapid response of telescopes upon alerts of transient objects is needed.

At the start of the 10-year LSST survey, the event broker must be capable of distinguishing known/unknown transients from the alert stream. The LSST project expects $10^6$ alerts per night (where an alert is defined as a 5σ difference from a reference image). This dwarfs the available spectroscopic resources. Many of these events will not require spectroscopic follow-up, and certainly most will not require immediate spectroscopic observation. But some objects with short lifetimes will need to be observed rapidly. The broker will have to winnow the alert stream down to a feasible number, on the order of one hundred objects. Without a software infrastructure to sort the alerts, the opportunity to thoroughly study rare and unusual events discovered by LSST will be lost.

Rapid (same day, or even same hour) alerts need to be issued for faint and fast objects such as .Ia SN, Ca-rich transients, GRBs and LIGO targets. The approximately ~100 objects/night that the broker deems interesting will need low-resolution (R~100-500) spectra with instruments such as FLOYDS[1] or the SED[2] machine on 2m (to reach r ~20 mag objects) and 4m (to reach r~22 mag) telescopes. The low-resolution spectra will provide more information for further characterization of the alerts. This will narrow the optimum target list of the most interesting/unusual objects to about 10/night.

A change of observatory operation modes over all apertures from 2-30m will be needed to enable spectra on demand, which could be dedicated facilities (optimum), queue scheduling or TOO interrupts.

In order to understand the nature of variability, obtain radial velocities for masses, determine metallicity, temperature and gravity, we will need moderate resolution (R~3000-5000) with wide wavelength coverage (the atmospheric cutoff to the K band) on 4-8m class telescopes of a few thousand selected targets per year down to g,r ~23 mag and about 100 per year on a 20-30m telescope for even fainter objects.

---

[1] http://lcogt.net/user/dsand/dev/Site/FLOYDS_Spectrograph.html
[2] https://sites.google.com/site/nickkonidaris/sed-machine

The cadence of these spectra will depend on the object, with spectra obtained over days-months to follow the development of SN and Novae, over days for GRBs, over weeks for accretion disk systems and over months for AGN. Over the course of the survey, about 10000 spectra of SN, 200 novae, dozens of accretion disk transitions and a few AGN will be needed to accomplish science on those objects.

Long-term spectral follow-up (after 1-5 years of the survey when light curves of particular variable objects are complete) will be needed to determine stellar population statistics, stellar evolution, oddities of various classes of variables, parameters like metallicity that vary across our Galaxy and the differences between galaxies. Once the frequency and distribution of various targets is determined, the costs/benefits of multi-object vs. single-object or small field IFU can be determined. The number of spectra will depend on the specific type of target, e.g., several thousand for Cepheids, several hundred for the end products of close binary evolution. Moderate resolution on 4-8m targets can be used for RV work down to r~22 mag for emission line objects, 20-30m to reach objects in other galaxies.

2.2 Galactic Structure (Knut Olsen facilitator)

The discovery space enabled by LSST for Galactic structure and stellar populations is big, with the science making particular use of the all-sky coverage, the large depth of the coadded survey, the capability to detect large numbers of variables as tracers of structure and populations (e.g. RR Lyrae out to 400 kpc), and the ability of LSST to measure parallaxes and proper motions at faint magnitudes. And while it is true that the discovery potential of LSST by itself is tremendous, in almost all cases it is also easy to see how spectroscopy provides enormous added benefit, e.g., through the addition of line-of-sight velocities, measurement of stellar abundances, or other detailed information on individual sources.

The group's discussion on the need for spectroscopy to follow up LSST discoveries began by identifying several broad themes of particular importance to our scientific interests. The group then narrowed the discussion to two that appeared to drive the spectroscopic requirements most strongly: 1) Galactic structure and 2) the stellar populations of the solar neighborhood. Specific science cases are described in the subsections that follow. In summary, the group identified compelling cases for several kinds of spectrographs:

- Massively multiplexed optical spectrographs, on a range of aperture sizes, including 4-m, 8-10-m, and larger telescopes, covering a range of resolution from R~2000 – 20,000.
- Moderately multiplexed and/or single-object optical spectrographs on 8-10-m or larger telescopes, with resolutions R~10,000 and higher
- High-resolution single-object near infrared spectrographs, with R~40,000 – 50,000, on 8-10-m or larger telescopes

- Low-resolution single-object near infrared spectrographs, with R~2,000, on 4-m or larger telescopes

The group also identified the potential utility of targeted narrowband imaging follow-up as a complement to spectroscopy, which points to the need for further consideration of the imaging and photometric follow-up needs in the era of LSST. Finally, we noted that many of the Galactic structure themes discussed by the group echo those found in the Feasibility Study Report for the Next Generation CFHT (http://orca.phys.uvic.ca/~pcote/Feasibility_Science_Final.pdf).

2.2.1 Galactic structure

LSST enables an enormous discovery space for the study of Galactic structure. From photometry of the coadded images, LSST will be able to use old main sequence turnoff (MSTO) stars as probes of structure out to ~200 kpc, or ~4× further (~16× larger volume) than for planned DECam surveys; the long lifetimes of MSTO stars make them useful for tracing structures with equivalent surface brightness of ≥35 mag arcsec$^{-2}$. LSST will be able to follow RR Lyrae out ~400 kpc, probing the structure of the intergalactic region between the Milky Way and M31. We identified several specific science questions in the area of Galactic structure that would particularly benefit from spectroscopy.

2.2.1.1 What is the accretion history of the Galaxy?

As described by Freeman & Bland-Hawthorn (2002, ARA&A, 40, 487) in their review article *The New Galaxy*, we are entering an age where we anticipate having the ability to piece together the early accretion history of the Galactic halo and thick disk by identifying the stars that remain of the fragments from which these components formed. LSST by itself will be an excellent tool for identifying spatial structures, like the now well-known Sagittarius Stream, that are the signatures of recent Galactic satellite accretion events. But stars left over from such events become mixed throughout the Galaxy through dynamical interaction over several Gyr (e.g. Helmi & White 1999, MNRAS, 307, 495), requiring additional phase-space information to identify the fragments from which they came. Using giants, horizontal branch stars, subgiants, and MSTO stars as potential tracers, spectroscopy of large samples of stars can add measurements of line-of-sight velocities, bulk metallicities ([Fe/H]), alpha-element abundances ([$\alpha$/H]), carbon abundances ([C/Fe]), and abundances of key individual elements. When combined with proper motions from GAIA and LSST, we will be able to construct the full dynamical and chemical phase space of the stellar samples and begin to identify individual early stellar systems. The characteristic mass of the fragments that we will be able to identify will be limited by the sample size; with a sample of $10^7$ thick disk stars and $10^6$ halo stars, we would typically have ~10 stars each from a population of fragments with characteristic mass of $10^4$ M$_{sun}$. With smaller samples, we would only be sensitive to the more massive fragments.

The clear need for this science case is multiobject spectroscopy, on a range of aperture sizes. The typical observing depth will be r~22 and fainter, probing a volume of 40 kpc radius with MSTO stars, 100 kpc with HB stars, and 400 kpc with RGB stars. The S/N needs range from <10 for measurement of velocities with <10 km s$^{-1}$ accuracy, to ~20 for measurements of [Fe/H], [$\alpha$/H], and [C/Fe], ~30 for individual $\alpha$-elements, and 50-100 for a range of other elements. Resolution needs also vary, with R~2000-5000 adequate for velocities and the coarser abundance measurements, and R~20,000 or higher for the most detailed abundance work. Most of the lines of interest are in the optical wavelength range.

Several questions important to the science case were not immediately answered by the group, including:

- How hot a stellar tracer can we effectively use? Measurements of abundances are more difficult at higher temperatures as spectral lines become fewer and non-LTE effects creep in, while velocities become less accurate with broader lines.
- What is the sweet spot for the velocity error? More accurate velocities will give finer dynamical resolution (though at some point random motions in the populations themselves make higher accuracy unnecessary), while coarser velocity measurements can make use of lower S/N spectra and poorer resolution, yielding larger samples.
- How do we efficiently select targets? The science case targets the thick disk and halo components specifically, but these account for <~10% of the Galaxy's stellar mass. Efficient selection is key for making the best use of telescope time.
- How many different samples are needed? We are unlikely to be able to collect every type of measurement (velocities and abundances) for every star in the sample, because of the range of brightness, resolution, and S/N requirements.
- How accurately can we measure the photometry? High photometric accuracy translates directly to improved target selection efficiency and accuracy of abundances and distance estimates.

2.2.1.2 What is the shape of the dark matter halo?

The shape of the dark matter halo (its radial density profile and angular dependence) is a key ingredient to any model of the formation of the Galaxy. Measuring the DM halo shape requires a large number of stellar tracers distributed all over the sky, with velocity measurements with accuracy <10 km s$^{-1}$ and, where available, proper motions for constructing 3-D orbits. Red giants with r~22 would allow us to probe the DM halo out to ~400 kpc.

The spectrograph requirements for this science case are essentially identical to that of 2.2.1.1. Because the goal is to only measure velocities, most of the work could be done through multiobject spectroscopy on 4-m-class telescopes. Resolution needs are R~2000-5000. This science project could easily piggyback on the more demanding observations needed for studying the accretion history of the Galaxy.

2.2.1.3 What does the population of DM subhalos look like?

In a hierarchical structure formation scenario, the Galactic halo is expected to be populated by hundreds of lower mass dark matter subhalos. While the number of dwarf satellites known to reside in the Galactic halo continues to rise as new systems are discovered, there remain far fewer dwarf satellites than would be predicted if all DM subhalos hosted stellar systems. If a large population of missing dark satellites exists, the only way to detect them will be to observe their gravitational effect on visible matter. A promising method, enabled by LSST's precise and deep photometry, is to identify cold streams in the halo and look for velocity perturbations from dark subhalos nearby. The most useful streams are the coldest ones, such as the tidal tails around the globular cluster Palomar 5, which have a velocity dispersion of ~2 km s$^{-1}$ (Odenkirchen et al. 2009, AJ, 137, 3378). The project would use the deep LSST photometry to identify well-defined stellar streams with potential substructure (e.g. Ibata et al. 2002, MNRAS, 332, 915; Johnston et al. 2002, ApJ, 570, 656; Carlberg 2009, ApJL, 705, 223), and select ~1000 targets per stream for spectroscopy. In order to maximize the number of targets, we will use the base of the RGB, subgiant stars, and MSTO stars as necessary, for typical magnitudes of r~20.

The most fundamental spectrograph requirement is to be able to measure velocities with precision <1 km s$^{-1}$, which demands R~10000 or better. The relatively faint magnitudes that are likely needed and required medium to large sample sizes would be best served by multiobject spectroscopy on 8-10-m class telescopes. Because we would be targeting relatively narrow streams, the field of view of these spectrographs need not be particularly large, although ~1 degree field of view would be helpful in improving the targeting efficiency.

Questions for exploration:
- How many streams are needed in order to have a likely probability of detecting at least one DM subhalo?
- Given the mass spectrum and distances of likely detected streams, what is the magnitude distribution of targets given a requirement of ~1000 sources per stream?

2.2.1.4 How long is the metal-poor tail of the Galactic halo?

The most metal poor stars in the Galactic halo provide a valuable window into the early history of the Universe, as their chemical abundance patterns reflect the products of at most a few generations of chemical enrichment.  Finding the stars that occupy the extreme of the metal-poor tail could provide a view of the very first generations of supernovae, perhaps even those produced by the first Population III stars.  The challenge in finding them is to identify these very rare stars from the background sea of much more numerous metal-rich stars.  LSST colors will provide the rough first cut, while a massive low-resolution spectroscopic survey on a mid-sized telescope would be needed to provide a second cut.  Once promising candidates are identified, they would be followed by high-resolution high S/N observations on large telescopes.  The target stars need not be faint, and as such represent potential low-hanging fruit.

For selecting the most promising candidates, low resolution (R<2000) spectroscopy with S/N~20 is needed, of as large a sample of stars as possible, beginning with the brightest (r<~17) stars.  The first stage thus calls for highly multiplexed spectroscopy on 4-m-class telescope.  This demand for low resolution spectroscopy could be significantly reduced by efficient use of narrowband imaging targeting the Calcium H&K lines and the G band.  Once the candidates are identified, single object, high-resolution (R>20000) spectroscopy on 8-10-m class telescopes is needed to study the abundances of individual elements with high precision.

2.2.2 The Solar Neighborhood

LSST's precise astrometry over a 10-year baseline across the full southern sky makes it a powerful tool for constructing a complete inventory of the solar neighborhood out to ~200 pc through parallax measurements.  While GAIA's regime will be to provide astrometry for stars brighter than r~20, LSST will provide far superior performance for the faint end of the white dwarf cooling sequence and substellar dwarfs.  The huge volume within which we can study brown dwarfs with LSST will, when combined with spectroscopy, allow us to address some exciting fundamental questions about them.

2.2.2.1 What are the masses of brown dwarfs?

Direct knowledge of stellar masses across a broad range of temperatures is fundamental to modeling and understanding their characteristics.  For brown dwarfs, the number of objects with masses measured directly from their gravitational influence is very small.  The large number of brown dwarfs that LSST will discover within the solar neighborhood will turn up many in binary systems, where we will have the opportunity to measure their masses directly.  Measuring masses in binary systems requires repeat spectroscopic measurements in order to associate the velocity curves of the binary components with their orbits.  Spectroscopic measurements in eclipsing binary systems will be particularly valuable for measuring stellar parameters.

Because of their low effective temperatures, brown dwarf spectra peak in the near-infrared. Given their low masses, we need velocity accuracy of 50-100 m s$^{-1}$ to detect the velocity wiggles, corresponding to a near-infrared spectrograph with R~40,000 – 50,000. In order to achieve S/N~10 for single sources with K~15, the spectrograph will need to be on an 8-10-m or larger telescope.

2.2.2.2 What is the nature of weather on brown dwarfs?

The cool temperatures of brown dwarf atmospheres lead to complex atmospheric chemistry and behavior, including weather. Studying and understanding such phenomena is important both for modeling brown dwarfs and for understanding the physics of planetary atmospheres. LSST will discover a large number of brown dwarfs through parallax, as well as provide measurements of the photometric variability associated with atmospheric changes. Spectroscopy of the brown dwarfs, linked to the photometric variability, would be invaluable for understanding the phenomena.

Low-resolution (R~1000), near infrared spectroscopy with S/N≥20 is sufficient for monitoring the broad molecular and dust-cloud generated spectral features in brown dwarf atmospheres. With typical source magnitudes of K~15, the spectroscopy needs 4-m telescopes or larger.

2.3 Galaxies and AGN (Xiaohui Fan facilitator)

The overarching scientific motivation for the Galaxies and AGN breakout session was understanding the assembly, star formation, and chemical enrichment histories of galaxies, the interaction of galaxies with the IGM, the co-evolution of SMBHs and their feedback, all as a function of evolving LSS environment & cosmic time. The following science cases illustrate the capabilities desired to pursue these topics.

2.3.1 A Massive Redshift Survey

This would encompass a large-scale sampling of a range of environments and redshifts for chemical evolution, star formation rate, mass, etc. It would need several SDSS volumes for different redshift slices with approximately one million galaxies with diagnostic quality spectroscopy. This would entail coverage of an area of 1-3 degrees in diameter with ~1000 fibers and the spectral resolution to split [O II] (velocity sigma of 30 km/s, i.e., R~4000). Coverage of the faintest redshift bins requires 10-m class telescopes with adequate S/N for physical diagnostics (100s of nights on 10m class telescopes).

2.3.2 Topology of Reionization Survey (ToRS)

Reionization ended at z ≈ 6.5 and was probably due to photo-ionization from galaxies. It might well have been inhomogeneous, due to clustering of ionizing sources. LSST will detect and map the angular distribution of UV-bright galaxies out to z ≈ 7, with the main (wide) survey reaching brighter than L* galaxies at z=6-7 and fainter than L* at z ≤ 5. The deep-drilling fields reach fainter than L* at all redshifts. Spectroscopic follow up of LSST imaging will give an approximate map of the 3D UV luminosity density. This spectroscopy will measure the 3D distribution of the brighter galaxies and can be correlated with much denser sampling of LSST photometric samples to fainter magnitudes (fainter galaxies almost certainly dominate the emission of ionizing photons). Measurements from z ≈ 7 to z ≈ 3-4 will map the evolution of clustering and its relation to UV emission. This reaches lower luminosities and much higher space/surface densities at lower redshifts and permits extrapolation backward into the reionization era.

The ToRS survey will require optical (≤ 1μm) spectra to m=26-27, with the goal of detecting Lyα, Ly-break, and ISM UV absorption lines for measuring redshifts. Spectroscopic resolution of R~1000 would be adequate, but higher resolution would be better to improve the efficiency of detecting lines through the OH sky forest. A secondary goal will be to measure the Lyα emission fraction (which is sensitive to the neutral IGM) vs. the rest-frame UV luminosity. Multislit spectrographs would be favored to match to the high source density at very faint magnitudes (surface densities of ~few/arcmin2 at z ≈ 6-7, ~10/arcmin2 at z < 5) while retaining high throughput. The survey will require ~several deg² in order to sample the scale of reionization bubbles, which is roughly 1 degree (150 Mpc co-moving) at z=7. This project would use ~50 nights with GMACS on GMT.

2.3.3 The Regions of Highest Overdensity

LSST will find many (~tens) of regions of high overdensity, particularly at higher redshifts. These are unique environments for early assembly of galaxies and strong inflows. To study these regions, one would need coverage of ~10 arcminute diameter areas to measure 10-100 objects per region, with high enough spectral resolution to split the [O II] line pair (velocity error <30 km/s, i.e., R~400). Coverage of the highest z bins to 26th mag requires a 20-m class telescope. The spectra need to have S/N adequate for physical diagnostics and thus tens of nights on 10m class telescopes.

2.3.4 Ly Alpha Blobs

What are the Lyman Alpha blobs and what phase do they represent in the formation of galaxies, groups, and clusters? To study this will require identification of a large sample of LABs at z~2-5. We expect <1 per square degree. The spectroscopic requirements for identification include single object spectroscopy over 3200Å–1μm with high-throughput in the blue being essential. This would require ~100-200 nights on an 8m class telescope. Follow-up science would require near-IR spectroscopy (to get rest-frame optical line diagnostics, where possible) with single object slit spectroscopy acceptable, but IFU preferred. In addition, optical IFU spectroscopy would provide spatially resolved line ratios and kinematics. It would also be useful to have a small field (10 arcmin - 1 degree) follow-up multi-object spectroscopy to understand the environment, and possibly 2D spectropolarimetry. To calibrate photo-z's for the LBG population, one would want MOS optical spectroscopy of samples of ~100s to 1000s of LBGs for redshifts in the range of the LAB redshifts (z~2-5).

2.3.5 Dwarf Satellite Galaxies

LCDM predicts far more dwarfs than observed in Local Group, but we still have incomplete understanding of faint dwarfs beyond the Local Group. The challenge is in separating observationally those true dwarfs within the virial radius from 'background,' slightly more luminous galaxies. There are tens of parent galaxies at distances of 10-15 Mpc, with tens of faint dwarfs per host to be culled from 10,000 faint galaxies per square degree. For spectroscopic study, one would need coverage of 1-3 degree diameter with ~1000 fibers and accurate RV's for r~24 (R=2000). Ten thousand background objects per host is based on photo-z. This many sources would require multiple configurations of fiber MOS. For additional follow-up spectroscopy to get physical properties of true dwarfs for internal velocity dispersion and abundances, one would need R>8000.

2.3.6 IGM Tomography

Background AGNs and distant galaxies discovered by Ly alpha emission or Ly break energy distributions can provide high density probes for the 3-D structure of inflowing and outflowing gas from galaxies in the foreground, as well as its association with individual objects. To study this problem, one would need 10-arcmin FOV, MOS coverage of i=25.5 sources with S/N~10, hundreds of objects with accurate velocities (R~2000), and coverage from 0.4-1.0μm for rest-frame UV observations at 2<z<5. This would require 20 hours per field over tens of fields to cover a range of overdensities and redshifts.

2.3.7 Quasar Redshift Surveys

One of the least well-known aspects of the AGN (bolometric) luminosity function is the contribution of obscured AGN and the low-luminosity end of the unobscured population. The former extracts nuclear information from galaxy redshift surveys and multi-wavelength association of objects from other surveys. Unobscured objects are derived from the faintest objects captured by the LSST depth. The large sample allows determination of evolution in clustering, distribution of Eddington ratios, and relation of BH growth to galaxy growth. This builds on MS-DESI and current surveys. This requires a wavelength coverage of 0.38-1.26 um, R~1000-2000, and S/N~10 for velocity widths. A limit of i=24 gives 500 per square degree over 10,000 square degrees.

2.3.8 Reverberation Mapping

The response of the broad-line region to continuum variations creates relation of luminosity and line width allowing mass determination for the central BH. To date, this has been accomplished only for low-luminosity AGNs and then strongly extrapolated to higher luminosity AGNs. There is a possibility this might be a useful distance indicator for cosmology. There are two approaches, either an opportunistic trigger for strong variability or regular monitoring of known AGN in the deep-drilling field(s). This requires single object coverage for an all-sky trigger, MOS coverage of -1.5 deg diameter (deep-drilling), ~1000 fibers, and accurate RV's for r~24 (R>1000). There would be monitoring spectroscopy for the deep-drilling field(s) and new sequences for strong variables at first trigger.

2.3.9 Rare Classes of AGN

Special, astrophysically interesting classes of quasars/AGN will turn up given the LSST area and depth. For example, quasars at z>6 test early BH growth and are reionization/IGM probes. There will be ~100 z>7.2 quasars at Y<24 over the LSST survey. For these targets, an X-shooter-type instrument (0.8-2.5 micron), R~2000, single target, high S/N, on 10m class telescope would serve. For z>~6 quasars, at Y<24 there will be one object every 1-3 per square degree, or ~1 per PFS FOV. These could be observed as part of wide-area galaxy/quasar spectroscopic survey with R~2000 and moderate-to-high S/N on 6-10m class telescopes. Other examples include rare BALs to study feedback, quasar strong lenses, and ultraluminous high-z galaxies. To find these, one needs to develop target selection before/during commissioning with LSST-type filter/depth/ cadence.

2.3.10 Key Capabilities

Considering the above science case, the chief capabilities are highly multiplexed spectroscopic surveys and rare-object follow up. For the high-multiplex spectroscopic surveys, one would like 6.5-10m class telescopes with wide-field high-multiplexing optical/IR spectrographs of low-to-moderate resolution (R<=4000). These would be dedicated surveys with millions of targets. IR capability (realistically up to J band) is important for many science cases. Except for reverberation mapping, follow-up is not time sensitive. MS-DESI and PFS will provide much of this capability, but there is a noticeable mismatch between the currently planned facilities and the LSST footprint.

For rare objects and structure follow-up, the necessary capabilities are a 20m class telescope with a single/multislit/IFU spectrograph and a FOV <~10 arcmin. The resolution could be low-to-moderate (R<=4000) for most applications, but R~30000 for IGM abundance. IR capability (ideally with continuous coverage such as X-shooter) is important for many science cases. UV coverage is important for IGM tomography. Note that there is no X-shooter type instrument among first-generation instruments on the ELTs.

Other useful capabilities include an efficient IFU for high spatial resolution and rapid follow-up single object/IFU with broad wavelength coverage. In addition, a broker for transient alerts to identify possible AGN, especially flaring AGN, is important, along with other software development for target selection and image processing. A precursor survey with DECam and a complementary IR survey would help to facilitate LSST follow-up.

2.4 Dark Energy and Cosmology (Michael Wood-Vasey facilitator)

The Weak Lensing and Large-Scale Structure group identified several compelling science cases with different spectroscopic needs:

A) A survey of >100,000 bright objects (galaxies + QSOs) over >100 sq. deg. from $0<z<3.5$ for high-precision cross-correlation calibration of photo-z's; $R\sim4000$, 0.4–1.0μm
B) A survey of ~20,000–100,000 faint ($22 < i < 25$) galaxies from $0<z<3.5$ with highest possible redshift success rate for photometric redshift training, spanning wide area; $R\sim4000$, 0.4–2.0μm. This would be supplemented by observations of $i\sim25$ AB mag galaxies that do not yield secure redshifts at first pass, including:
- Longer exposures for objects that failed to yield redshifts
- JWST NIRSPEC or WFIRST IFU spectra of ~1,000 of the galaxies
- Spectroscopy using OH suppression technology could potentially yield significant benefits, although those have not been realized as of current generation instruments.

C) A cluster cosmology survey involving:
- 2500 spectra of red galaxies in clusters. The spectroscopic targets would be 2-3 galaxies from each of 1000 clusters in evenly distributed bins from $0<z<1.5$, with resolution $R\sim4000$ and coverage from 0.4–1.5μm for 100 km/s accuracy.
- Redshifts for >200 objects per cluster in 10 galaxy cluster candidates between $0<z<1.5$ for modified galaxy tests. The spectroscopy would have resolution $R\sim4000$ and cover 0.4–1.5μm.

D) A strong lensing cosmology survey using adaptive optics-corrected IFU spectroscopy for ~1000 strong lens systems on 20/30m class telescopes. The spectrograph would have $R\sim2000$, wavelength coverage 1–2 μm, a 5" field of view, and 0.05" sampling.

E) An SN Ia cosmology survey comprised of 10,000 SN Ia spectra at wavelengths 0.4–1.0μm with $R\sim100-1000$, which would be time-sensitive; and 200,000 SN host galaxy spectra, also at wavelengths 0.4–1.0μm but with $R\sim4000$, which would not be time-sensitive.

2.4.1 Photometric Redshifts to enable Dark Energy and Cosmology Studies with LSST

As LSST is a photometric survey, photometric redshifts and classification will impact all major extragalactic science cases. Many of the topics of cosmological investigations with LSST require the inference of redshifts and types of objects based on the LSST imaging data. The robust and accurate calibration of these inferences requires spectroscopic observations of significant subsamples.

2.4.2 Cosmology with Galaxy Clustering and Weak Lensing

2.4.2.1 Weak Lensing Story

The weak gravitational lensing of distant galaxies depends on the distribution of matter in the Universe and so is sensitive to dark energy through its effect on the growth of structures. The accelerated expansion of the Universe that is caused by dark energy opposes the gravitational attraction that would otherwise lead to the increased clumping of dark matter structures. The magnitude of the weak lensing signal depends on distances to both the lens and the source. The efficiency of the lensing changes slowly with redshift and thus relatively wide redshift bins can be used to study the evolution of the weak lensing signal.

This means the accuracy of individual photometric redshifts are not very important, but the distribution of photometric redshift errors must be known to high precision. For weak lensing we require an accurate estimation of the distribution of true redshifts within each photometric redshift bin.

2.4.2.2 Galaxy Clustering/LSS Story

For large-scale structure measurements, understanding and quantifying the overall uncertainties in galaxy photometric redshift estimates will be important for measurement of galaxy-galaxy lensing. Baryon acoustic oscillations (BAO) imprint a standard ruler that can be measured from the galaxy two-point correlation function. This measures the angular diameter distance and the Hubble parameter. BAO measurements will be performed by splitting the galaxy sample one by one into a series of redshift shells. This requires photoredshifts of the individual objects accurate enough to enable clean separation of bins. Overlap in redshift shells resulting from uncertainties will lead to cross-bin correlations.

2.4.2.3 Challenges

There are two main challenges. First, obtaining a highly accurate calibration of photometric redshifts ("Calibration"). The mean redshift and redshift spread (sigma) of LSST samples must each be known to $<\sim 0.003(1+z)$ for weak lensing and BAO dark energy inference not to be degraded. Second, producing the highest-quality photometric redshifts for every object ("Training"). Minimizing the uncertainty in the measurement of the photometric-redshift of an individual galaxy will have significant scientific benefits for BAO and cluster studies of dark energy, as well as galaxy clustering / environment studies. The SRD Requirement is that LSST should deliver sigma_z < 0.02(1+z) for perfect template knowledge. An optimal training set should bring us as close to this limit as feasible.

For Calibration, the requirements for LSS and weak lensing science with LSST can be met using cross-correlation calibration techniques that rely on using a spectroscopic sample that spans the redshift range of and overlaps in sky coverage with a photometric sample. While the LSST gold sample of galaxies will go down to i<25.3 AB mag out to z~3.5, the spectroscopic cross-correlation sample necessary to characterize redshift distributions (both mean redshifts and standard deviations) at the level required for dark energy inference need only span the redshift range, but not necessarily properties, of the gold sample. Thus redshifts for the most massive galaxies and most luminous quasars at a given redshift are sufficient. Because robust redshift identification is key, sufficient resolution to split the [O II] doublet is necessary. This implies 100,000 bright objects (r < 23.5) from 0<z<3.5 with R>4000 over 0.4μm-1.0μm.

For Training, the accuracy of clustering measurements (including BAO dark energy constraints), cluster identification, and constraints on the impact of intrinsic alignments on weak lensing will all be improved as the errors in individual-object photometric redshifts get smaller. The LSST system is being designed to be capable of delivering a gold sample per-object photometric redshift accuracy (with perfect knowledge of SED templates) of $\sigma_z < 0.02(1+z)$. The closer we can come to an ideal training set, the closer we will be to achieving this error level for all objects. One of the major concerns is obtaining secure redshifts for a fair sample of galaxies for all SED types and redshifts. >1% error rates in the calibration redshift sample would yield unacceptable systematic errors in photo-z calibration. Existing surveys to i~22.5 have achieved secure redshifts for 40%-70% of targets, with some failures broadly distributed in color space and others concentrated in limited regions. Galaxy types that are not represented in the redshift training sample cannot be reliably used for many types of scientific inquiry. An ideal training set would include a set of statistically complete (i.e. not systematically incomplete for any class of galaxies), secure spectroscopic redshifts for 20,000–100,000 faint objects i < 25.3 from 0<z<3.5. Maximal wavelength coverage is desirable; 0.4-1 um would be the minimum useful, while deep coverage from 0.37-2.0μm would be ideal. A minimum resolution of R~4000 will be needed to split the [O II] 3727A doublet, greatly enhancing the set of galaxies with secure redshifts. This also applies to the SN Ia host galaxies. Of order 1000–5000 objects may require the equivalent of JWST NIRSPEC spectroscopy to yield redshifts. Significant advances in OH suppression in multi-object spectroscopy would yield significant gains for ground-based efforts in support of these goals. An ideal, complete training sample would allow LSST calibration needs to be met without relying on cross-correlation techniques.

2.4.2.4 Observing Mode

For each of these photometric-redshift calibration needs, large survey programs will be appropriate and necessary to complete and enable these projects. These surveys could last as long as 5 or even 10 years; because it is feasible to get i<25.3 photometry over >100 square degrees with facilities now being commissioned, these efforts can begin well before LSST is on-sky.

2.4.3 Type Ia Supernova Cosmology

Ideal and perfectly accurate and calibrated SNe Ia could measure relative distances in the Universe to 0.07% per 0.1 redshift bin (7% /sqrt(10,000)).  However, we are currently limited by the systematics of the calibration ($\Delta w \sim 0.08$), the treatment of supernova color and dust ($\Delta w \sim 0.05$), and concerns of potential evolution of SN Ia properties with redshift ($\Delta w$ unknown).

The current known limiting systematic is calibration of the photometric systems. LSST will have a self-calibrated sample of SNe Ia ranging from 0.1<z<1.0.  The LSST calibration to 1% will reduce the $\Delta w$ contribution to ($\Delta w \sim 0.02$).  The investigation of color and dust will require significant UV, optical, and NIR imaging and spectroscopy in the years up to 2022.  A potential outcome of these studies is that SNe Ia will have to be matched up by galaxy properties to better control for the likely dust encountered by the SN Ia light in the host galaxy as well as an overall correlation with progenitor metallicity and other properties.  We are therefore most concerned about potential systematics that have not as yet been revealed.

Variations in intrinsic supernova properties likely come from their binary evolution and metallicity.  The metallicity, and potentially binary fraction/separation IMF are likely to be functions of the stellar formation history where the supernova is formed. A central concern is that these distributions and properties may evolve with redshift.  But the gross nature of this concern should similarly be well matched by the gross tracing of the properties of the host galaxy.  We thus strongly desire spectroscopy of the host galaxy of every supernova of all types (Ia, core-collapse, etc.). This will help solidify photometric-classification/redshift of the supernovae, and provide a way to map the properties of supernovae as a function of redshift.

The large number of LSST SNe Ia will allow for selection of ideal subsamples.  The selection of these subsamples will quite likely require information about the host galaxy.  Metallicity of the galaxy (both gas-phase and stellar) will likely be a key tracer, along with estimate of rates of recent star formation (last 200 Myr).

Type Ia supernovae cosmology will thus significantly benefit from spectra of ~10,000 supernovae at given phases.  These spectra will be distributed from 0.1<z<0.9 with 19 < i < 24.  This would be subdivided into three sets.  Set A would have 5,000 for spectra at maximum light.  Set B would have 2,500 spectra to follow all SN-like events to explore contamination and classification. Set C would have 2,500 spectra (5x500) repeated coverage to map out full coverage around peak.

R~100 is sufficient for supernova classification as these explosive events have broad features, with effective velocity widths from 3,000-10,000 km/s. It's possible that the ability to pick up narrower emission lines and absorption lines may be relevant for studies of SN Ia environments. But having R~100 capabilities on large apertures may be very helpful in efficiently obtaining spectra of z~0.9 SNe Ia. IFU observations would be preferable if possible to simultaneously observe the supernova and host galaxy.

In addition, spectra of 100,000 host galaxies from 0.1<z<0.9 would be useful. The main driver for R~4000 for the host galaxies is to obtain sufficient information on emission and absorption lines to characterize the gas-phase metallicity of the host galaxies and to split the [O II] doublet.

The specific observing capabilities to pursue the SN Ia science program include a transient broker to identify SN targets. For single SNe, one would like a single-slit or IFU, high-throughput, 0.4-1.2µm, R~500–1000 spectrograph on a 8-10m class telescope at the start of LSST operations. This would be a long-term program where targets would be allocated a few days ahead of time. There is no requirement for interrupts, but there is the need to be able to choose when a given target is observed. Understanding the selection effects introduced in target selection is critical to the success of contamination/rate/typing studies. This needs to be algorithmic and deterministic. There will be ~5 targets / square degree at a time. No significant multiplex advantage is likely possible, but coordination with large galaxy studies in the same area may yield gains. For the hosts of SNe Ia, the necessary capabilities are a multi-object spectrograph with wide (few degrees) FoV, 0.4-1.0µm, R~4,000. Observations could wait until well into the survey. A five-year survey could obtain all of the host galaxy redshifts with 100-500 targets per square degree after the full 10-year LSST survey.

2.4.4 Cosmology with galaxy clusters

Cluster cosmology relies on photometric redshifts to identify concentrations of galaxies. LSST is an exquisite photo-z + lensing machine, ideally suited to absolute mass calibration. It will find many optical clusters at z<1.2. Primarily spectroscopic redshifts would aid photo-z accuracy and enable the best lensing calibration. A second use will study ``clean,'' well-settled clusters to test gravity.

The primary tool for the absolute calibration of galaxy cluster masses for cosmological studies is weak gravitational lensing. Work in this area is progressing rapidly (e.g. von der Linden et al. 2013; Applegate et al. 2013) and it is likely that within the next 2-3 years the systematics in mass calibration with this method will become dominated by uncertainties in the photo-zs for galaxies in the cluster fields.

Most of the galaxies that enter cluster weak lensing analyses are relatively faint, i~25 mag. We currently have an incomplete picture of how well photo-z codes trained on field populations describe faint cluster members in this regime: do the faint cluster members have similar colors to field galaxies? Do we need to adjust the photo-z priors for cluster fields to reflect the greater likelihood of finding faint galaxies at the cluster redshift? The best way to address these questions is through comprehensive, deep spectroscopy of cluster fields over a range of cluster redshifts and also, ideally, cluster masses.

At a minimum, this work requires comprehensive spectroscopy of ~20 cluster fields (2 mass bins, 4 redshift bins, with 2-3 clusters per bin). Using KECK, or similar 8-10m class facilities, we expect to attain good (~75%) completeness down to r~24 with 1 (ideally 2) nights per cluster. To reach r~25, will require 30m telescopes and again 1 (ideally 2) nights per cluster. In total, this implies ~20 (ideally 40) nights of both 10m and 30m telescope time. Other possible instruments to initiate this work at brighter magnitudes, with the ability to efficiently target large numbers of objects, include the Low-Dispersion Prism at Magellan, as well as HST Grism surveys (e.g. the GLASS survey).

2.4.5 Cosmology with strong lensing

Strong lensing constraints on cosmology rely on having kinematic mass measurements for hosts, as well as highly accurate position measurements for both the host and images. High-spatial resolution IFU spectroscopy can fill both these needs. We expect to identify ~1000 'gold' high-quality lensing systems with LSST.

The ideal capability for this purpose would be high-strehl adaptive optic IFU spectroscopy for ~1000 strong lens systems using 20/30m class telescopes with R~3000 and 1-2 um coverage. The required field of view is the typical size of time-delay lenses, 3-4" on a side. A secondary need is for spectroscopy of the more massive objects over ~few arcmin radius from the lensing system to constrain line of sight convergence and shear. Kinematic information will be helpful for establishing masses. An ideal sample would constitute 100 objects/lens system, observed with spectral resolution R > 3000, over ~5 arcmin radius field of view, spanning 0.4-1 um wavelength coverage or broader.

**3. Synthesis**

There are several common capabilities across the various topics discussed at the workshop. This included not just instruments, but observing modes and software infrastructure.

3.1 Software Infrastructure

Although time domain science is the primary beneficiary of a broker to sort through the alerts generated by changes in brightness (or position), the cosmological use of

SNe Ia, the study of AGNs, and the study of brown dwarf weather all will rely on the ability to recognize interesting events when they happen.  For each case, the signal hidden in the noise of one million alerts per night will be different, but they will all require a system to winnow alerts done to the ones they will find of interest.  Hidden among those million alerts will be rare and interesting objects, and these will be lost without a working broker.

3.2 Instruments

3.2.1 Low-Resolution Spectrographs

Although a broker will be important for narrowing down the number of alerts to a reasonable number, there will be some events that will require more information in order to make a decision.  For truly transient events, when an object appears where none has before, LSST will provide only a magnitude at discovery.  There will be no history to help characterize the event.  For these objects, a rapid, low-resolution spectrum can provide more information.  Because of their low resolution (R~100-500) and high throughput, they can be deployed on smaller aperture facilities (2-4m).  A suite of such spectrographs on a range of telescope apertures would greatly enhance the identification of rare and interesting objects in the LSST alert stream.  It would also help to identify SNe Ia for cosmological use, AGNs for variability and reverberation mapping.

3.2.2 Single-Object, High-Throughput, Wide-Wavelength Coverage Spectrographs

For many objects, the density on the sky at any given time will be low enough that there is no advantage to a multi-object spectrograph.  While this will apply mainly to time-domain events, it is also true of rare non-variable objects (such as high-redshift AGN and galaxies).  The general requirement for resolution is to split the [O II] doublet, so R~3500, but this is also high enough to separate the OH night sky lines and thus produce better spectra of faint objects in the infra-red.  Most time-domain targets will be unknown and the known targets will cover a wide redshift range, so the widest possible wavelength coverage is important.  This covers from the atmospheric cutoff at 0.32µm to the K band.  A single instrument that encompasses all these capabilities may be difficult, but a suite of instruments deployed on a range of telescope apertures from 4m to 10m would fulfill most of the need.

3.2.3 Highly Multiplexed Spectrographs

There are many science cases from across the breakout sessions that require spectra of many thousands of objects.  Some would like orders of magnitude more.  The instruments would be wide field, with a field of view of 1-3 degrees.  There would be capacity for hundreds (minimal) to thousands (highly desirable) of spectra in a single observation.  These cases focus mainly on optical wavelengths, but IR capability is also important.

For the multi-object cases, there is less commonality when it comes to resolution. Most extragalactic programs and some Galactic structure science could use low-to-moderate resolutions of R~1000-3000. The Galactic science programs also desire higher resolutions of R~20,000-30,000. These would be deployed on telescopes with apertures of 4m to 10m, with the larger telescopes taking most of the demand.

3.2.2 Single-Object High-Resolution Optical and Near-IR Spectrographs

For Galactic science, availability of high-resolution optical and near-IR spectrographs on the largest telescopes is very important. High-resolution optical spectrographs, with R~20,000 and higher, are particularly important for measuring the properties of the rare extremely metal-poor stars in the Galactic halo, from which we will have a unique window into the early Universe. High-resolution near-IR spectrographs, with R~40,000–50,000, are critical for deriving masses of ultracool dwarfs in low mass binary systems.

3.3 Facilities

Building new facilities is an expensive proposition, so repurposing existing facilities may be the answer to meeting the need for spectroscopic follow-up in the era of LSST. From all the science cases, it is clear that there is a need for a range of apertures. Smaller apertures from 2-4m can help with classification and study of the brighter end of LSST discoveries. The 6.5-10m telescopes are more suited to the follow-up of the bulk of the targets found by LSST.

In addition, the operation of these facilities may need to change. Dedicated operations that are matched to specific science goals will more directly enable follow-up of LSST targets. This way, scheduling and time allocation can be adapted to the most efficient use of these facilities. It may be easier to dedicate smaller aperture facilities, but the volume and rate of LSST targets suggests that dedicated larger facilities would also be valuable.

3.4 Timing

LSST is scheduled to begin operations in 2022, so design and construction of some instruments has to start now. The survey, though, will last for ten years. Given the length of the project, not every follow-up capability has to exist at the start of operations. Rare transients are unlikely to be uniformly distributed over the survey, so the capabilities to discern and study them must be in place once LSST begins producing alerts.

For static objects, the follow-up can occur on a more leisurely time-scale. The massively multiplexed spectrographs can be on sky well into the period of LSST operations. In fact, it may take several iterations of data-catalog release in order to produce the samples that are of interest, so there may be little or no scientific gain in having these capabilities early in LSST operations.

As the various science goals of the LSST project have different operational timelines for follow-up, this allows for the development of a strategic plan for capability design and construction. In addition, this spreads the cost of these capabilities over a longer time, allowing for a more consistent approach to funding.

## 4. Acknowledgments

We wish to thank the participants in this workshop (listed in Appendix 2) for their valuable contributions, as well as contributions from indviduals who could not attend, including Steve Allen and Rachel Mandelbaum. We would also like to thank the members of the organizing committee, Todd Boroson, Steve Heathcote, Ken Hinkle, Joan Najita, and Steve Ridgway.

# Appendix 1. Spectroscopy Summary Table

This table represents an initial attempt to characterize parameters for individual projects that will follow up LSST discoveries. It is not meant to be exhaustive, but illustrative. It can be the framework that guides future discussion.

| Problem[a] | Depth[b] | S/N[c] | $\lambda$[d] | R[e] | $\Sigma_{Target}$[f] | Area[g] | Comments |
|---|---|---|---|---|---|---|---|
| Superluminous SNe | $16 < r < 25$ | $> 5$ | $0.4 - 2.5\mu m$ | 2000 | $0.05\ deg^{-2}$ | entire LSST footprint | |
| Cataclysmic variables | $16 < r < 25$ | $> 5$ | $0.4 - 2.5\mu m$ | 2000 | $10\ deg^{-2}$ | entire LSST footprint | |
| Galaxy stellar dynamics | $16 < r < 25$ | $5 - 10$ | $0.4 - 0.9\mu m$ | 2000–5000 | ... | $18,000\ deg^2$ | |
| Galaxy stellar abundances: [Fe/H], [$\alpha$/Fe], [C/Fe] | $16 < r < 25$ | 20 | $0.37 - 0.9\mu m$ | 2000 | ... | $18,000\ deg^2$ | |
| individual $\alpha$ elements | $16 < r < 25$ | 30–50 | $0.37 - 0.9\mu m$ | 5000 | ... | $18,000\ deg^2$ | |
| "all" individual elements | $16 < r < 25$ | 50–100 | $0.37 - 0.9\mu m$ | 20,000+ | ... | $18,000\ deg^2$ | |
| Brown dwarf masses | $K \sim 15$ | $> 10$ | $1.0 - 1.6\mu m$ | 5,000 | ... | $18,000\ deg^2$ | |
| Massive galaxy survey | $K \sim 15$ | $> 10$ | $1.0 - 1.6\mu m$ | 5,000 | ... | $18,000\ deg^2$ | |
| Brown dwarf weather | $20 < i < 25$ | $> 5$ | $0.4 - 1.3\mu m$ | 4000 | $1000\ deg^{-2}$ | $1000\ deg^2$ | |
| Topology of reionization survey | $z_{AB} \sim 26 - 27$ | $> 5$ (for Ly$\alpha$ line) | $5000 - 1\mu m$ | $1000 - 4000$ | up to 10 arcmin$^{-2}$ | $1\ deg^2$ | |
| Dwarf satellite galaxies | $r < 24$ | continuum $> 10$ for RV | $4000 - 9000$Å | 4000 | $10,000\ deg^{-2}$ | few deg$^2$ | |
| IGM tomography | $i < 25 - 26$ | continuum S/N $\sim 10$ | $3500 - 10000$ Å | 2000 | $10\ arcmin^{-2}$ | FOV 10 arcmin | |
| Quasar redshift survey | $r < 24$ | 5 | $3800 - 12600$ | $1000 - 2000$ | $500\ deg^{-2}$ | $10,000\ deg^2$ | |
| Reverberation mapping | $i < 24$ | 10 | $4000 - 10000$ | $> 1000$ | $1000\ deg^{-2}$ | one field (1-2 deg$^2$) | |
| $z > 6$ quasars (other rare AGN) | $Y < 24$ | 5 | $0.8 - 2.5\ \mu m$ | $> 2000$ | single object | entire LSST footprint | |
| Ly$\alpha$ blobs | $i < 24$ | 5 | $3200 - 6000$ Å | 2000 | single object | entire LSST footprint | |
| Weak Lensing/LSS cross-corr. cal. | $20 < i < 23$ | ... | $0.4 - 1.0\mu m$ | 4000 | $1000\ deg^{-2}$ | $100\ deg^2$ | |
| Weak Lensing/LSS photo-z train. | $22 < i < 25$ | ... | $0.4 - 2.0\mu m$ | 4000 | $1000\ deg^{-2}$ | $5000 - 10000\ deg^2$ | |
| Weak Lensing/LSS supplemental | $22 < i < 25$ | ... | $0.4 - 2.0\mu m$ | 4000 | $10\ deg^{-2}$ | $5000 - 10000\ deg^2$ | |
| Cluster Cosmology photo-z cal. | $i < 25$ | 5 | $0.4 - 1.5\mu m$ | 4000 | $100\ deg^{-2}$ | $100\ deg^2$ | |
| Strong Lensing cosmology | ... | ... | $1 - 2\mu m$ | 2000 | $1/10\ deg^{-2}$ | $10000\ deg^2$ | |
| SNIa Cosmology: SN follow-up | $gri \sim 19 - 24$ mag | ... | $0.4 - 1.0\mu m$ | 1000 | $5\ deg^{-2}$ | $20000\ deg^2$ | |
| SNIa Cosmology: Host follow-up | $20 < i < 25$ mag | ... | $0.4 - 1.0\mu m$ | 4000 | $30\ deg^{-2}$ | $20000\ deg^2$ | |

| Problem | Min. Sample Size[h] | Desired Sample Size[i] | Target Efficiency[j] | # Visits[k] | Cadence[l] | When[m] | Comments |
|---|---|---|---|---|---|---|---|
| Superluminous SNe | few 100 | few 1000 | 1 | several | 3-4 days | start of LSST | X-shooter-like |
| Cataclysmic variables | $10^4$ | $10^5$ | 1 | several | variable | start of LSST | near UV; near-IR ; IFU |
| Galaxy stellar dynamics | $10^6$ | $10^7$ | ... | 1 | ... | Throughout | Any bright objects |
| Galaxy stellar abundances: [Fe/H], [$\alpha$/Fe], [C/Fe] | $10^6$ | $10^7$ | ... | 1 | ... | Throughout | faint galaxies |
| individual $\alpha$ elements | $10^6$ | $10^7$ | ... | 1 | ... | Throughout | Likely JWST or WFIRST+IFU |
| "all" individual elements | subset | subset | ... | 1 | ... | Throughout | |
| Brown dwarf masses | ... | ... | ... | 1 | ... | Throughout | |
| Massive galaxy survey | ... | ... | ... | 1 | ... | Throughout | |
| Brown dwarf weather | a few million | 2000 | 1 | $\sim 100$ | deep drilling | static | |
| Topology of reionization survey | 1000 | 10,000 | 0.5 | 1 | co-eval with LSST | static | |
| Dwarf satellite galaxies | 10 fields | few tens of fields | 1 | 1 | ... | static | needs to be done for a number of redshift bins |
| IGM tomography | 10 fields | few tens of fields | ... | 1 | ... | static | for several different redshift bins |
| Quasar redshift survey | $10^5$ | 100,000 | ... | 1 | ... | static | higher R for velocity dispersion |
| Reverberation mapping | 500 | 2000 | ... | 1 | ... | start of LSST | |
| $z > 6$ quasars (other rare AGN) | 1000 | 10,000 | ... | 1 | ... | static | |
| Ly$\alpha$ blobs | 100 | 500 | ... | 1 | ... | static | |
| Weak Lensing/LSS cross-corr. cal. | 50,000 | 100,000 | ... | 1 | ... | static | |
| Weak Lensing/LSS photo-z train. | 20,000 | 100,000 | ... | 1 | ... | static | |
| Weak Lensing/LSS supplemental | 500 | 1000 | ... | 1 | ... | static | |
| Cluster Cosmology photo-z cal. | 1000 | 2500 | ... | 1 | ... | static | |
| Strong Lensing cosmology | 500 | 1000 | ... | 1 | selective | static | IFU: 0.05"/samp. over 5"x 5"; prob. 20m/30m |
| SNIa Cosmology: SN follow-up | 5,000 | 10,000 | ... | 1 | selective | start of LSST | |
| SNIa Cosmology: Host follow-up | 100,000 | 200,000 | ... | 1 | static | static | |

[a] Particular astronomical question or source.  [b] Brightness ranges of targets.  [c] Desired signal-to-noise ratio.  [d] Desired wavelength coverage.  [e] Resolution.
[f] Target density (multi- vs. single-object spectroscopy, background).  [g] Necessary survey area.  [h] Minimum number of targets.  [i] Desired number of targets.  [j] Desired target selection efficiency (purity).  [k] Number of visits per target.
[l] Desired cadence.  [m] When is capability needed relative to the start of LSST operations.

**Appendix 2. Workshop Participants**

- Alexandra Abate (Physic Dept/University of Arizona)
- Felipe Barrientos (P. Universidad Catolica de Chile)
- Timothy Beers (NOAO)
- Fuyan Bian (Steward Observatory/University of Arizona)
- Robert Blum (NOAO)
- Todd Boroson (NOAO)
- Zheng Cai (University of Arizona)
- Elliott Cheu (University of Arizona)
- Chris Copperwheat (Liverpool John Moores University)
- Katia Cunha (Observatorio Nacional/Steward Observatory)
- Darren DePoy (Texas A&M University)
- Arjun Dey (NOAO)
- Mark Dickinson (NOAO)
- Xiaohui Fan (University of Arizona)
- Federico Fraschetti (University of Arizona)
- Brenda Frye (Steward Observatory)
- Richard Green (Steward Observatory)
- Kenneth Hinkle (NOAO)
- Željko Ivezić (University of Washington)
- George Jacoby (GMTO/ Carnegie)
- Buell Jannuzi (Steward Observatory)
- Nick Konidaris (Caltech)
- Tod Lauer (NOAO)
- Young Sun Lee (New Mexico State University)
- Steve Margheim (Gemini Observatory)
- Jennifer Marshall (Texas A&M University)
- Rachel Mason (Gemini Observatory)
- Thomas Matheson (NOAO)
- Alan McConnachie (NRC Herzberg Institute of Astrophysics)
- Ian McGreer (University of Arizona)
- Bryan Miller (Gemini Observatory)
- Joan Najita (NOAO)
- Jeffrey Newman (U. Pittsburgh)
- Dara Norman (NOAO)
- Knut Olsen (NOAO)
- Vinicius Placco (IAG-USP/NOAO)
- Stephen Pompea (NOAO)
- Stephen Ridgway (NOAO)
- Susan Ridgway (NOAO)
- Abhijit Saha (NOAO)
- David Sand (Texas Tech University)
- Edward Schmidt (University of Nebraska-Lincoln)
- Sam Schmidt (UC Davis)
- Rob Seaman (NOAO)

- Nigel Sharp (NSF)
- Richard Shaw (NOAO)
- David Silva (NOAO)
- Anze Slosar (BNL)
- Chris Smith (NOAO/CTIO/AURA)
- Verne Smith (NOAO)
- Letizia Stanghellini (NOAO)
- Michael Strauss (Princeton University)
- Paula Szkody (University of Washington)
- Rene Walterbos (NMSU)
- Benjamin Weiner (Steward Observatory)
- Grant Williams (MMT Observatory)
- Michael Wood-Vasey (University of Pittsburgh)
- Dennis Zaritsky (Steward Obs./Univ. of Arizona)